\documentclass[letterpaper]{article}
\usepackage{aaai2026} 
\usepackage{times} 
\usepackage{helvet}  
\usepackage{courier}  
\usepackage[hyphens]{url}
\usepackage{graphicx}
\usepackage{amsmath}
\usepackage{natbib}  
\usepackage{caption} 
\usepackage{booktabs}
\usepackage{subfigure}
\usepackage[inline]{enumitem} 
\usepackage{nameref}  
\usepackage[dvipsnames]{xcolor}
\makeatletter
\gdef\showauthors@on{T}
\makeatother

\usepackage{newfloat}
\usepackage{listings}
\DeclareCaptionStyle{ruled}{labelfont=normalfont,labelsep=colon,strut=off} 
\title{Triangulating Across U.S. Federal AI Transparency Regimes}
\author{Emma Lurie\textsuperscript{\rm 1}, Emma Fauser\textsuperscript{\rm 2}, Qing He\textsuperscript{\rm 1}, Danaé Metaxa\textsuperscript{\rm 1}, and Sorelle A. Friedler\textsuperscript{\rm 2}}
\affiliations{
    \textsuperscript{\rm 1}University of Pennsylvania, Department of Computer and Information Science, Philadelphia, PA, USA\\
    \textsuperscript{\rm 2}Haverford College, Department of Computer Science, Haverford, PA, USA\\
    ewlurie@seas.upenn.edu, efauser@haverford.edu, qingh@design.upenn.edu, metaxa@seas.upenn.edu,  sorelle@cs.haverford.edu\\

}

\newcommand{\website}{\url{https://govAIfiles.org}} 
\newcommand{\github}{\url{https://github.com/triphora/govaifiles.org/}}

\begin{document}

\maketitle

\begin{abstract}
Federal AI systems can deny benefits or flag individuals for deportation, but the public disclosures meant to make those systems visible are fragmented and unevenly detailed. This paper examines three existing U.S. federal transparency regimes---System of Records Notices (SORNs), Information Collection Requests (ICRs), and the AI Use Case Inventory---and asks how well they, individually and together, describe government AI use. We find that no single regime fully reveals how the government constructs or deploys AI: each discloses different aspects of a system, and the current disclosure infrastructure makes it very challenging for the public to track specific AI systems across regulatory regimes and over time. Persistent identifiers are absent, granularity varies widely, and the annual AI Use Case Inventory cycle means federal agencies can deploy systems months before appearing in any official record. Using hand-validated zero-shot classification and cross-document entity resolution, we contribute a triangulation method that links disclosures across all three regimes and present two case studies. Our case studies finds that linking records provides greater insight into government AI use, but even linked records would constitute insufficient oversight compared to what public reporting has revealed about the same systems. We trace each regime's disclosure weaknesses to its original administrative purpose, showing these gaps are structural, and offer recommendations focused on the AI Use Case Inventory as the mechanism best suited for public-facing transparency: (1) a broad and consistently applied AI system definition, (2) persistent system identifiers with cross-references to related disclosures, and (3) restored public visibility into risk management processes.\end{abstract}

\begin{links}
    \link{Website}{https://govAIfiles.org}
    \link{Code and data}{https://github.com/triphora/govaifiles.org}
\end{links}

\section{Introduction}

In December of 2025, a U.S. citizen was walking home when he was stopped and detained by U.S. federal immigration officials. Agents handcuffed him and scanned his face using a phone app; he was only released when the app confirmed his citizenship. Reporting revealed that agents likely used Mobile Fortify, a facial recognition and fingerprint-matching app used by Immigration and Customs Enforcement (ICE) 
to determine a subject's citizenship status.

The app, first reported by 404 Media \citep{cox2025ice}, raised a number of questions and concerns about AI governance and accountability in a U.S. federal context: was this app being used as directed, were there avenues for redress and oversight, and was its use even legal?  Proposals for answering these and other AI governance concerns often rely on transparency mechanisms \citep{ostp2022blueprint, oecd2019ai, kingsman2022public}. But new transparency mechanisms may or may not be needed; the U.S. federal government already operates under disclosure requirements, including some designed specifically for AI and others 
that may disclose government AI use related to databases and data collection. 

In this paper, we identify three existing U.S. federal government transparency regimes (further detailed in Section~\ref{sec:existing}) and consider the extent to which they make AI applications 
visible to the public. These existing regimes are:

\begin{enumerate}
\item \textbf{System of Records Notices (SORNs)} are required notices whenever a federal agency creates or changes the functionality of a database containing personal information. Designed in the 1970s with a focus on government data storage, SORNs may now reveal information about AI systems that make use of that data. 
\item \textbf{Information Collection Requests (ICRs)} comes out of the Paperwork Reduction Act and requires agencies to report on any data collection processes, whether data is collected by paper or electronically; when agencies collect data for an AI system, this may be revealed via ICRs and supplementary documents such as Privacy Impact Assessments (PIAs). 
\item Finally, the federal \textbf{AI Use Case Inventory} is a recent effort specifically tailored to make AI use by the federal government transparent. All Executive Branch agencies are required by law to disclose all AI systems that are in use by their agency to the Office of Management and Budget (OMB) and disclosed to the public. 
\end{enumerate}

\textbf{Our research considers what existing federal AI disclosure regimes reveal and obscure about government AI use, and what cross-regime comparisons illuminate that single-regime analyses cannot.} Linking records across regimes is a substantial methodological challenge: the three regimes share no common system identifiers, were not designed to be interoperable, and each uses vocabulary shaped by its own administrative logic. Using zero-shot learning and cross-document entity resolution techniques developed to bridge disparate text corpora, we link AI deployments across all three transparency regimes. 

We find that the triangulation of AI regulatory regimes reveals insights that no individual transparency regime provides. Using two high-impact systems as case studies---\textit{Automated Targeting System} and \textit{Mobile Fortify}---we demonstrate that connecting multiple transparency regimes allows stakeholders to better understand the assemblage of systems, data sources, models, and oversight that comprise the entirety of the system. We find that it is currently difficult for the public to track a specific AI system across regulatory disclosures and over time: persistent identifiers are absent, between 80 and 97 percent of records in each corpus have no confirmed cross-regime link, and the annual AI Use Case inventory cycle means systems can be deployed for months before appearing in any official record. Further linkage and transparency concerns include vocabulary divergence, inconsistency in the amount of desired and revealed detail, and mission creep, such that a system use changes without requiring the associated disclosure to be updated. 

We close with policy recommendations for improving government AI transparency. The AI Use Case Inventory is a promising mechanism, but several fixes would improve its efficacy towards public transparency: removing the ``principal basis'' language that limits the AI definition, adopting persistent system identifiers with cross-references to other government filings, and restoring public visibility into risk management processes. We can reform the existing disclosure ecosystem without whole cloth reinvention.

\subsection{Contributions}
\noindent In summary, this paper contributes:
\begin{enumerate}
    
    \item \textbf{Evidence that triangulation adds value}: using \textit{Automated Targeting System} and \textit{Mobile Fortify} as case studies, we show that cross-regime analysis reveals more about high-impact AI systems than any single transparency regime currently can.
    
    \item \textbf{A characterization of systemic gaps in federal AI transparency}: we identify three recurring failure modes --- timing mismatches, inconsistent granularity, and mission creep --- and trace each to the history and administrative logic of the underlying mechanism.

    \item \textbf{A public transparency tool for navigating the AI Use Case Inventory}: We develop \website\, a user-friendly web application that makes it easier for the public to search and explore government AI use.
    
    \item \textbf{A cross-regime linking methodology}: we develop a zero-shot learning and cross-document entity resolution method to connect AI disclosures across three federal transparency regimes not designed to interoperate.
    
    \item \textbf{Policy recommendations for improving federal AI transparency}: we offer targeted recommendations for the AI Use Case Inventory, including clarification to the definition of covered AI systems, the adoption of persistent system identifiers, and reinstating robust risk management disclosures.
\end{enumerate}

\section{Literature Review}
\subsection{The role of transparency in AI governance}

Transparency is widely recognized as foundational to AI governance \cite{jobin2019global}. When government entities deploy AI systems, individuals can rarely opt out of being subject to those systems, making transparency an important mechanism for correcting the informational and power asymmetries between governments and constituents \cite{hickok2024public}. Transparency is a prerequisite for rights-preserving AI governance: without the ability to know how, why, and to what end AI systems are deployed, other accountability mechanisms are undermined \cite{algorithmwatch2020automatic}. 

Yet disclosure alone does not guarantee meaningful transparency. Ananny and Crawford's \cite{ananny2018seeing} concept of “seeing without knowing" explains that having knowledge of a system, is not the same as knowing how it works or being able to govern it. Often, efforts at various forms of transparency or explainability have actually been found to be directed at assessing specific properties of a system, such as its potential for discrimination, contestation, or recourse \cite{selbst2018intuitive, ustun2019actionable}. Algorithmic impact assessments are a common operationalization that assesses the achievement of normative goals, such as non-discrimination, via transparent reports that document the process and results of system tests \cite{reisman2018algorithmic} and that have been proposed as part of governance processes in the vein of environmental impact assessments \cite{selbst2017disparate}. Algorithmic impact assessments form an increasing part of governance approaches to AI \cite{selbst2021institutional}, and there are many methodological concerns about how best to use these instruments as effective governing tools \cite{watkins2021governing} especially with regards to including marginalized communities in the determination of what impacts are measured \cite{metcalf2021algorithmic}. The AI use case inventory we study brings some of these concerns to the fore---impact assessments are required as part of the risk management disclosures for high-impact AI systems.

\subsection{Studies of Government AI Use}
The public record on how artificial intelligence systems are built, funded, and overseen by the U.S. federal government is scattered across government websites and difficult to piece together into a coherent narrative. Gaps or fragmentation in disclosure translate directly into gaps in oversight, leaving affected communities with limited recourse.
Scholars have approached this problem by examining different government datasets. Wang et al. \citeyearpar{wang2022american} show that ICE constructed a domestic surveillance infrastructure through procurement relationships that largely bypassed public scrutiny, drawing on data from FOIA requests and procurement records that had previously generated little public attention. Levendowski \citeyearpar{levendowski2021trademarks} locates transparency in federal trademark filings. Because the USPTO requires evidence of use, trademark registrations can compel private surveillance companies to disclose operational details about products that other regulatory channels (FOIA, patents) tend to obscure. Bateyko and Levy \citeyearpar{bateyko2025one} identify a different overlooked site of AI governance: federal grant notices. Agencies directing billions in discretionary funding shape how state and local entities develop and deploy AI, yet this grantmaking activity has developed largely outside the transparency norms that govern procurement. Across these three bodies of work, a pattern holds: consequential AI governance activity occurs in administrative registers that were not designed with AI transparency as a goal, and are rarely analyzed together.

What remains underexplored is whether existing federal disclosure regimes, examined in combination, can yield meaningful public insight into high-impact AI systems, and what cross-regime comparison reveals that any single regime cannot. We examine three such mechanisms (System of Records Notices, Information Collection Requests, and the AI Use Case Inventory), tracing how each renders AI systems visible or invisible, and where their coverage diverges. The aim is not to call for new regulatory infrastructure wholesale, but to identify targeted reforms grounded in the administrative history and logics of what already exists.

\section{Existing Transparency Regimes and AI}
\label{sec:existing}

Debates about government surveillance, administrative burden, and public accountability for automated systems and data collection are not unique to the AI era \cite{ruggles2023s}. They have produced a layered set of disclosure regimes stretching back to at least the 1970s \cite{hew1973records}. In this section, we identify these existing regimes and describe how they could help us gain visibility into AI system use.

\subsection{System of Records Notices}
System of Records Notices emerged from 1970s concerns about government computerization and its implications for individual privacy. In drafting the Privacy Act of 1974, Congress relied on an advisory committee report documenting the risks presented by the increasing use of electronic information technologies, which had begun replacing traditional paper-based government systems \cite{hew1973records}. To address these risks, the report recommended five Fair Information Practice Principles: openness, individual access, collection limitation, use and disclosure limitations, and information management \cite{hew1973records}. These principles remain at the core of privacy legislation at the federal, state, and international level \cite{gellman2025fip, hartzog2017inadequate}.

The Privacy Act operationalized these principles by requiring federal agencies to publish a System of Records Notice in the Federal Register whenever an agency creates or modifies a system containing personal information about individuals. 
These notices must include the system's name and location, the types of individuals and records included, how the information will be routinely used and by whom, the agency's policies on data storage and retention, and procedures for individuals to access and challenge their records \cite{privacy_act_1974}. Thus, AI systems can be revealed when included as part of the description of how the data will be used, and when identified as the data underlying an AI system, the description of the data can shed light on how such systems work.

\subsection{Information Collection Requests}
The Paperwork Reduction Act (PRA) regulates information-collection burdens imposed by the federal government. Enacted in 1980 and revised in 1986 and 1995, it requires agencies to obtain OIRA approval before collecting information from ten or more members of the public~\cite{shapiro2012pra,shapiro2020reinvigorating}. The review process requires agencies to justify the need for the collection, estimate the burden imposed on citizens, provide opportunities for public comment, and display a valid OMB control number \cite{shapiro2012pra,levy1994pra}. Notice of the collection must be filed with the Federal Register. In cases where personally identifiable information (PII) is collected, agencies are also required to file a SORN before filing their ICR \cite{OPM2011}.
OIRA's goal is to minimize federal information-collection burden while also maximizing the utility of the information collected. More broadly, OIRA acts an institutional ``information aggregator'' that collects dispersed information from agencies, the Executive Office, state and local governments, and the public, and as a ``guardian of a well-functioning administrative process'' \cite{sunstein2013oira}. It is this aggregation feature that we take advantage of in this study as we analyze the documentation included in Information Collection Requests (ICRs) and posted to a federal government website.

The goal of the PRA was to respond to concerns over the cost burdens, often in terms of time, imposed on citizens, state and local governments, and businesses by federal paperwork~\cite{levy1994pra,samaha2015death}. Federal agencies collect a large amount of information from the public and retain substantial discretion over the form and burden of that collection. The PRA's aim was to force agencies to screen out duplicative, poorly designed, or unnecessary collections \cite{levy1994pra}.
The statute's performance is often criticized. On aggregate burden reduction, there is limited evidence that the PRA has achieved either large-scale burden reduction or accurate burden estimation \cite{shapiro2012pra,shapiro2020reinvigorating,pahlka2023recoding,samaha2015death}. Yet, OIRA review improves some individual collections and may deter some ill-considered collections from ever being proposed \cite{shapiro2012pra,shapiro2020reinvigorating}. The PRA is therefore best understood as a procedural quality-control regime with uneven aggregate effects: it improves some collections and creates accountability but has not reliably reduced the cumulative paperwork burden it was designed to control.

Importantly for our purposes, agencies file an ICR with OIRA before manually or automatically collecting digital or biometric information as well as more traditional paperwork. The result is that AI systems built on such collected data, such as facial recognition systems, can be found described in these filings.

\subsection{AI Use Case Inventory}
Unlike the regimes that preceded it, the AI Use Case Inventory was designed explicitly around AI. The inventory was originally required under the first Trump Administration via an executive order on trustworthy AI (EO 13960), and later codified in law under the Advancing American AI Act. The Biden Administration continued this implementation, and provided associated guidance on risk management through an OMB management memorandum (M-24-10 \cite{omb_m2410}) which was updated by the second Trump Administration (M-25-21 \cite{omb_m2521}), though remained substantively the same in many ways \cite{friedler2025omb}.

Based on reporting guidance from OMB associated with these memoranda, agencies must annually report publicly and to OMB's Office of the Federal CIO (OFCIO) each AI system used by the agency, whether developed in-house or acquired for use, with exceptions for research systems, national security systems, and a few other uses. Each reported use case must include basic information about that use case, such as its purpose and the type of AI system, as well as risk management information in the case of high-impact systems. High-impact systems are identified via a specific definition in the OMB management memoranda and have an associated list of specific systems that are presumed to be high-impact \cite{omb_m2410, omb_m2521}. These include risk assessment systems related to immigration, biometrics used in public spaces, vehicle tracking of non-governmental vehicles in a law enforcement context, patient diagnosis systems, and other immigration, law enforcement, healthcare, government services, and safety or rights related AI system uses.

\section{Methods for Triangulating Across Regimes}

In order to develop a dataset on which we can conduct a cross-regime analysis, we first collect and filter the data to include only identified AI-related entries and standardize the resulting information. Then we create methods for triangulating across these collected AI-related entries.

\begin{table*}[ht]
\centering
\caption{Dataset collection and AI-related findings by disclosure regime.}
\label{tab:dataset-overview}
\begin{tabular}{@{}lp{4cm}p{4cm}p{4.4cm}@{}}
\toprule
 & \textbf{AI Use Case Inventory} & \textbf{SORN} & \textbf{ICR} \\
\midrule
\textbf{Coverage} &
  FY2023, FY2024, FY2025 &
  2010--2025 (Federal Register) &
  Jan 2018--Mar 2026 (OIRA) \\
\addlinespace
\textbf{Collection unit} &
  Use case record, per agency &
  Published notice &
  Information collection request \\
\addlinespace
\textbf{Total collected} &
  710 (2023); 2,071 (2024); 3,542 (2025) &
  3,735 notices with full text & 37,871 ICRs; 31,670 with downloaded docs \\
\addlinespace
\textbf{AI identification method} &
  Agency reported &
  Zero-shot LLM classification &
  Two-stage keyword filter + zero-shot LLM \\
\addlinespace
\textbf{AI-related records} &
  All records (reported by agency) &
  554 of 3,735 (15\%) &
  1,146 of 3,378 classified ICRs (34\%)\textsuperscript{a} \\
\bottomrule
\end{tabular}
\vspace{4pt}
\footnotesize\textsuperscript{a}~The 34\% rate reflects only ICRs whose supporting documents passed the keyword pre-filter, not the full corpus.
\end{table*}

\subsection{Identifying and Standardizing AI-related Entries}

\subsubsection{SORNs}
We identified AI-relevant Systems of Records Notices (SORNs) using a zero-shot large language model classification workflow applied to a corpus of federal SORN JSON files. The underlying corpus consisted of 3,735 SORNs spanning 2010-2025 pulled from the Federal Register API. 
We then classified these SORNs as AI-related using zero-shot classification on GPT-5.4 mini. Model outputs were reviewed on a random sample of 122 SORNs flagged as AI-related by the classifier, of which 85 (69.7\%) were confirmed and 37 (30.3\%) were incorrectly labeled as AI-related. The final prompt, available in Appendix~\ref{app:sorn-prompt}, distinguished systems that perform matching, scoring, recognition, prediction, ranking, approval or denial support, risk assessment, or other automated analytical judgments from systems that primarily store, retrieve, exchange, or display records. It also returned a binary \texttt{is-ai-related} flag indicating whether machine learning or AI was explicitly stated in the notice, short quoted evidence phrases copied from the SORN text, and brief reasoning. Model outputs were then reviewed as part of the evaluation process.  

\subsubsection{ICRs}
We collected all Information Collection Requests (ICRs) received by the Office of Information and Regulatory Affairs (OIRA) from January 1, 2018 through March 31, 2026, yielding 37,871 ICR records. We scraped ICR listings and detail pages from \url{reginfo.gov}; each ICR's detail page was parsed to extract structured metadata including the submitting agency, ICR reference number, title, abstract, and links to supporting documents.

To identify AI-relevant materials within this corpus, we applied a filter to identify OMB supporting statements, Privacy Impact Assessment (PIA), Privacy Threshold Analysis (PTA), System of Records Notice (SORN), Algorithmic Impact Assessment (AIA), or civil liberties assessments; as well as documents that contained a predefined set of AI-related vocabulary (see Appendix~\ref{app:icr_processing} for the filter procedure and AI-related vocabulary). After filtering, 3,378 ICRs moved onto the classification stage.

Candidate documents were classified using the GPT-5.4-mini model. Each document was submitted with a structured prompt (see Appendix~\ref{app:pra-prompt}) asking the model to assess whether the described information collection involved AI. Classification was applied at the document level; ICRs with at least one AI-related document were flagged as AI-related at the ICR level. We manually reviewed 100 of the ICRs that were flagged as AI matches, and found 75\% to be AI-related. The full batch produced 1,146 AI-related ICRs (33.9\% of the 3,378 candidates). See Appendix~\ref{app:icr_processing} for further details.

\subsubsection{AI Use Case Inventory}
We analyzed all federal agency AI Use Case Inventory disclosures (2023-2025). To support cross-agency and cross-year comparison, we normalized column names, response categories, and standardized agency and bureau names to canonical forms using a crosswalk and fuzzy-matching procedure (see Appendix ~\ref{app:inventory_processing} for details). Because all entries in the AI Use Case Inventory are AI technologies, no classification was required.

\subsection{Cross Regime Analysis}
\label{sec:linkage}
A methodological challenge in this paper is determining when records from the three regulatory regimes (SORNs, ICRs, and AI Use Case Inventory) refer to the same underlying AI system. These documents comply with different regulatory requirements, focus on different elements of AI systems, use different vocabularies for important terms, are filed by different offices at different times, and rarely cross-reference one another.\footnote{While ICRs that involve PII require an agency to file a SORN first, they sometimes still may not include the associated SORN or SORN identifier directly in the ICR.} The task of linking them is an instance of \textit{cross-document entity resolution} \cite{beheshti2017systematic}: identifying when textually dissimilar records share a common real-world referent. The three regime-pairs (SORN--ICR, SORN--Inventory, ICR--Inventory) are treated as separate subtasks, because they differ substantially in difficulty: SORN--ICR pairs share regulatory context and tend to have stronger structural similarities, while ICR--Inventory pairs are often the most difficult because ICR filings describe data \textit{collections} while the Inventory describes \textit{systems}.

Following best practices \cite{jain2024knowledge, beheshti2017systematic, zhao2023cross}, we do the following steps: (1) candidate retrieval; (2) probablistic scoring; (3) LLM reranking; (4) clustering, and (5) evaluation. 

\subsubsection{Candidate Retrieval} 

The full cross-product of AI-relevant records (591 SORNs, 1,146 ICRs, and 4,558 AI Use Case Inventory entries) yields tens of millions of possible pairs across the three regime combinations; evaluating each individually is infeasible. Candidate retrieval reduces each pair space to a manageable set without discarding true links.
For each record, the pipeline retrieves the 20 most similar records in the paired regime using an additive retrieval score:
\begin{equation}
s_{\text{retrieval}} = 0.5 \cdot s_{\text{TF-IDF}} + 0.2 \cdot s_{\text{agency\_name}} + 0.3 \cdot s_{\text{named\_entity}}
\label{eq:retrieval}
\end{equation}

\noindent where $\text{tfidf}$ is the cosine similarity between TF-IDF vectors (shared vocabulary per pair, up to 10,000 unigram and bigram features, log-scaled term frequency to reduce the influence of document length); $\text{agency}$ is 1.0 if both records belong to the same parent agency and 0.0 otherwise; and $\text{ne}$ is a normalized count of shared system acronyms, vendor names, and program names extracted via regular expression and named-entity recognition.

Two design choices follow from the structure of the data. First, whether or not the same agency filed the request is additive rather than a hard filter: cross-agency pairs with high TF-IDF similarity still appear in the top-20 candidates, because early results found cross-agency adoption cases, where a system developed by one agency is later used by another, among the most analytically important findings. Second, date proximity is not included in the retrieval score; qualitative review of early candidates found it to be a poor signal for links, since systems frequently generate regulatory filings years apart.

\subsubsection{Probabilistic Scoring} Each candidate pair surviving retrieval is scored on a weighted combination of features: whether the records come from the same agency, whether they share system names or acronyms, whether they mention the same vendors or programs, and how semantically similar their descriptions are. Formally:
\begin{equation}
\begin{split}
s_{\text{composite}} =~ & 0.25 \cdot s_{\text{agency}} + 0.25 \cdot s_{\text{name}} + 0.30 \cdot s_{\text{named\_entity}} ~+ \\
& 0.15 \cdot s_{\text{semantic}} + 0.05 \cdot s_{\text{TF-IDF}}
\end{split}
\label{eq:prob_scoring}
\end{equation}
\noindent where 
$\text{name}$ is a fuzzy string similarity between system titles (exact match scores 1.0, shared acronyms score at least 0.8, otherwise token-set ratio); and $\text{semantic}$ is the cosine similarity between OpenAI \texttt{text-embedding-3-small} embeddings. The TF-IDF cosine is retained at low weight (0.05) as a weak residual signal. Semantic similarity receives a lower weight than the structural features because regulatory language varies substantially across regimes and embedding models may not reliably capture domain-specific paraphrase relationships.

Pairs are assigned to one of three bands: scores above 0.62 advance to the LLM reranker as high-confidence candidates; scores between 0.30 and 0.62 advance as borderline candidates; and scores below 0.30 are rejected as confident non-links. Threshold values are calibrated against 8 known explicit SORN--ICR cross-references confirmed by document cross-citation. 

\subsubsection{LLM Reranking} Pairs above the scoring threshold are submitted to a LLM for a classification with a structured explanation (see Appendix~\ref{app:reranker_prompt}). The model is asked to classify each pair as same system, related (operationally connected but legally distinct---e.g., a data collection filing that feeds a separately described AI model), or no link, and to cite specific phrases from each document that support its decision. The ``related'' category matters analytically: it captures dependency relationships between systems that should not be collapsed into a single record but are still meaningful for understanding how disclosure gaps propagate across regimes. 

\subsubsection{Clustering} Accepted same-system links from all three regime-pairs are aggregated into clusters using agglomerative clustering with group-average linkage, terminating when inter-cluster similarity falls below a threshold of 0.7 \cite{gao2024enhancing}. This resolves a single AI system that appears across multiple regimes 
into one cluster rather than a set of disconnected pairs.
A useful property of this approach is that cross-agency transitive links are recovered naturally: if a DHS system and a DOL system are each independently linked to a common AI Inventory entry, they are placed in the same cluster even if the DHS--DOL pair was never directly scored. Records classified as \textit{related} by the reranker are not merged into clusters; they are preserved as a separate set of data-dependency edges.

\subsubsection{Evaluation} Because no labeled dataset exists for cross-regime AI system linking, we construct an evaluation set through annotation.  We draw a stratified sample across three groups:  links (cross-regime, same system), proposed related systems (shared data dependencies),  non-links (pairs the system rejected with high confidence). One of the authors reviewed 100 random pairs of each of the  groups and finds accuracy rates of: (1) 85\% (same system); (2) 51\% (related system); and (3) 100\% (having no link).

\section{System: \website}
\label{sec:website}

We develop \website\ which contains a searchable web application and interactive data exploration features; each is described below. These components are available in a GitHub repository \github.

\begin{figure}[t]
    \centering
    \includegraphics[width=\columnwidth]{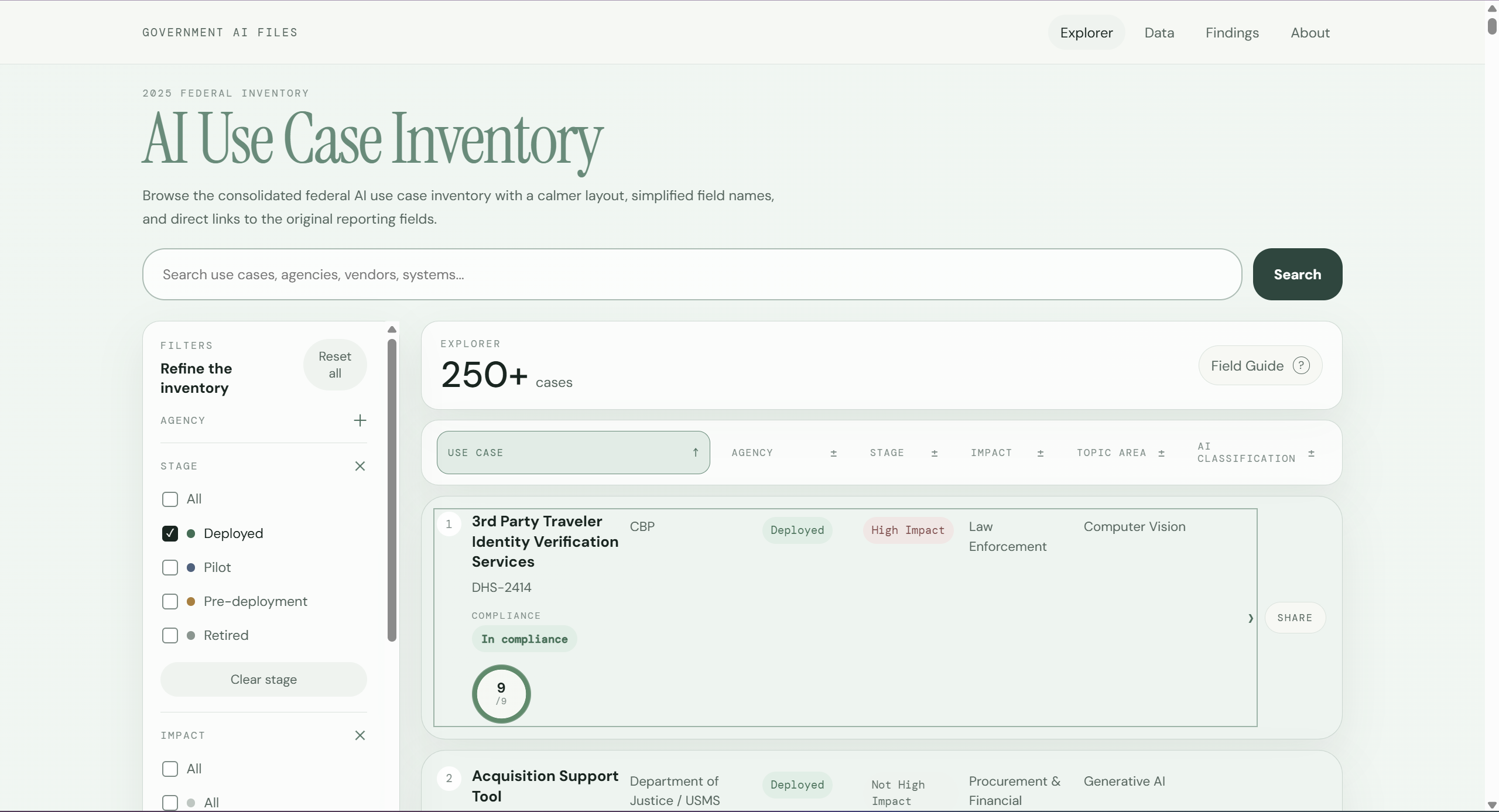}
    \caption{A screenshot of the home page of the \website\ search application.}
    \label{fig:homepage}
\end{figure}

\paragraph{Web application} We build a web application that allows searching and filtering of AI Use Case Inventory records (see Figure~\ref{fig:homepage}). The application sets up a single database table with the consolidated AI Use Case Inventory records and associated metadata such as year published. The application imports the database table into our own search engine that only contains these records, then serves an API interfacing with our search engine. The search engine accepts two optional inputs: (1) a single query string, which will cause the output to only contain results with a match for the query in any text field of the AI Use Case Inventory records, and (2) a list of filters (e.g., year, agency, deployment stage), which will cause the output to only contain results matching every filter provided. The API interface directly passes the inputted query and filters to the search engine, and is publicly available for use. Detailed breakdowns of each AI Use Case Inventory are then easily parsable by individuals seeking information about government transparency documents (see Appendix Figure~\ref{fig:mobile_fortify}).

\paragraph{Interactive data exploration.} We include in the web interface a bar graph with interactive options to show the quantities and attributes of AI Use Case Inventory records. 
The graph shows the quantity of Inventory records reported by year and can be filtered and grouped by stage of development, high impact status, and agency/bureau.

\begin{figure}
    \centering
    \includegraphics[width=\linewidth]{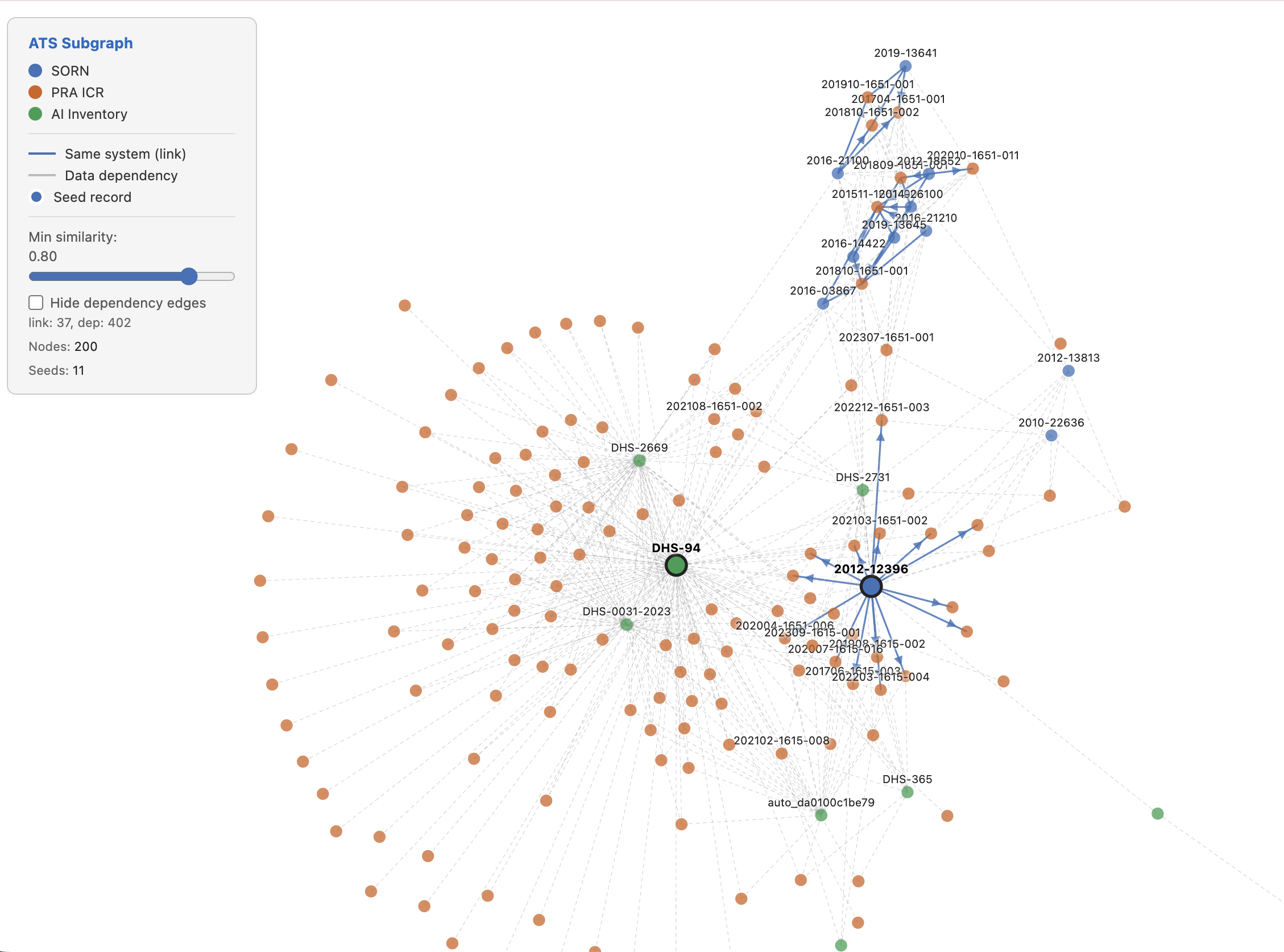}
    \caption{Nodes represent records from three federal disclosure regimes — SORNs (blue), ICRs (orange), and AI use case inventory entries (green). Solid edges connect records linked to the same underlying system across regimes; dashed edges indicate data dependencies, where one system draws from ATS. The graph reveals dense data-dependency structure relative to same-system links, alongside several tight clusters of SORNs and ICRs that document the same system under different disclosure obligations.}
    \label{fig:ats}
\end{figure}

\paragraph{Researcher backend.} To enable deeper investigation of AI systems across disclosure regimes, we provide a cross-regime record linkage pipeline that links entries in the AI Use Case Inventory to related SORNs and ICRs. Using the method described in Section~\ref{sec:linkage}, we grouped items into cross-regime.
The resulting link graph is available to researchers as structured outputs that can be queried by system, agency, or regime pair. To illustrate the depth of analysis this enables, we construct an ATS (Automated Targeting System) subgraph (see Figure~\ref{fig:ats}): seeded with AI inventory entries and one SORN for ATS, the pipeline surfaces almost 200 additional connected records across all three regimes, including ICRs that document data flows into ATS from immigration and customs forms.

\section{Comparing Across Regimes: A Richer Picture of Federal AI Use}
The developed cross-regime analysis methods allow us to create and identify clusters of interest.  We focus on two examples that emerged from that analysis. The first, the Automated Targeting System (ATS), is a decision-support tool for border crossings characterized by not only some direct linkages across ICRs and SORNs, but also a rich graph of data dependencies where ATS feeds into another data collection or analysis system. The second, Mobile Fortify, the facial-recognition app, demonstrates that even when there are few direct textual matches, a broader cross-regime analysis can provide more context about data sources and governance. 

\subsection{Case Study: Automated Targeting System (ATS)}
The Automated Targeting System (ATS) is CBP's decision-support tool for risk assessments for everyone who crosses a U.S. border. This case study highlights two findings. First, ATS exemplifies how bringing disclosures, fragmented across regimes, into conversation can reveal more information about government AI use. Second, ATS demonstrates the idea of mission creep, as it is a system, described in documentation with a narrow purpose, that has been repurposed and repackaged across uses and systems.

ATS emerged as one of the densest hubs in the clustering analysis, with connections spanning all three disclosure regimes (SORNs, ICRs, and AI Use Case Inventory) and a data-dependency structure that made it a natural candidate for closer examination (Figure~\ref{fig:ats} shows the subgraph). We define data dependencies as a link between systems where ATS is named as the processor or subject of data collection. The density of data dependency connections (dashed edges) relative to different parts of the same system (solid edges) is itself a finding: ATS is a frequent data source for AI-related systems across government.

ATS is a long-standing and controversial technology \cite{stanley2007acluATS}. Its governing Privacy Act notice (with SORN identifier DHS/CBP-006) was first published in 2006, describing a system that performed “rule based query of the cargo or personal information accessed by ATS" \cite{SORN2007ATS}. A 2012 update expanded the scope of data the system ingests, adding over dozen new source systems including the FBI's Terrorist Screening Database, but retained the same core vocabulary, describing ATS as ``rule-managed technology that facilitates the targeting of high-risk travelers and cargo performing database comparisons, scenario-based targeting, and query functions'' \cite{SORN2012ATS}. The notice has not been substantively updated since. Yet the 2025 AI Use Case Inventory discloses ten machine learning models operating within or feeding into ATS, none of which appear by name in the SORN. The SORN contains none of the words “artificial intelligence,” “machine learning," or "algorithm." Whatever machine learning related capabilities ATS has acquired by 2025, the governing notice vocabulary has never been updated to reflect them.

Linking ATS to the AI Use Case Inventory allows information not provided elsewhere to be identified. Four of the ten models are classified as high-impact: the Passenger Security Assessment Model, Cargo Security Assessment Model, Illicit Trade model, and Traveler Entity Resolution. Two of the four, the Passenger Security Assessment Model and Traveler Entity Resolution,  directly score individual travelers and draw on demographic variables including sex and age. All ten are categorized as Classical/Predictive Machine Learning or Computer Vision and designated as deployed systems. Their descriptions use language nowhere present in the Privacy Act notice: “advanced AI and machine learning models," “predictive modeling," ``risk scores." Of the ten, two list ManTech as vendor (Passport Anomaly, DHS-81; Trade Entity Risk, DHS-95); the remaining eight are built in-house.

Beyond the AI Use Case Inventory, 166 ICRs, overwhelmingly from CBP and USCIS, name ATS as the downstream processor of collected data, tracing a supply chain that feeds models the SORN does not acknowledge. ATS deployments extend beyond CBP and USCIS. For example, two CDC records, governing COVID-era collection of airline passenger contact information, explicitly disclose that the data will be processed by ``the DHS Automated Targeting System (ATS)." Several ICR supporting documents surface references to facial comparison technology operating within ATS, a capability the SORN never mentions (e.g., DHS/CBP/PIA-056, CBP/PIA-068). Moreover, a July 2017 ATS Privacy Impact Assessment (supplementing a 2022 ICR filing) partially fills in the post-SORN record, describing ATS as a modular decision-support system and acknowledging facial recognition and other AI components. 

The cross-document accounting necessary to reconstruct who developed ATS technology and how it is used points to a structural problem in how AI transparency is currently framed. The conventional unit of analysis for AI disclosure is the model: a bounded artifact with an owner and a clearly delineated purpose. ATS resists that framing. The SORN describes a data storage and query platform; the inventory names the models; the ICR records trace the data flow from individuals. No single document connects those layers, and the 166 records naming ATS as a downstream processor suggest that the more analytically useful unit may be the data supply chain rather than the system itself. A traveler's information may flow through a form collection, into ATS, and be scored by a model that the SORN does not acknowledge, with no single disclosure document that follows that path end to end.  That fragmentation is also what allows mission creep to go unaccounted for: a system first described in 2006 as performing ``rule based query" of traveler information can by 2025 be running ten deployed machine learning models, four of them high-impact, processing data from agencies as far afield as CDC, without any major updated privacy or risk analyses that reflect the change.

\subsection{Case Study: Mobile Fortify}
Mobile Fortify is a DHS mobile application that captures facial images, fingerprints, and photographs of identity documents to verify the identity of individuals. It transmits those inputs to CBP, which runs the biometric matching; Mobile Fortify displays the returned results to ICE users. CBP owns the matching infrastructure; ICE operates the front-end mobile application. Mobile Fortify illustrates two limits of federal AI transparency infrastructure: (1) documentation lag that can appear as absence, and (2) gaps in disclosure such that system-level understanding can remain incomplete even when cross-regime links are assembled.

The 2025 AI Use Case Inventory is the only transparency disclosure that names Mobile Fortify directly. It discloses two deployed, high-impact use cases: one for CBP and one for ICE. Both report that required pre-deployment safeguards (e.g., impact assessments, risk management protocols, and monitoring mechanisms) are still in progress. This potentially leaves it out of compliance with OMB's AI memorandum \cite{omb_m2521}.\footnote{Due to the annual cadence of the disclosures, it's possible it became in compliance before OMB's deadline for agencies in April 2026 and after the 2025 Inventory was released in January 2026.}

Mobile Fortify does not appear by name in either the SORN or ICR datasets. The two inventory records reference ``TVSI" and ``ATS-C" as associated systems, without definitions. ``TVSI" most likely refers to TVS-Internal, the internal-facing variant of CBP's Traveler Verification Service. ``ATS-C" is less obvious, plausibly a shorthand for ATS-Cargo, 
but no available document connects ATS-Cargo to Mobile Fortify’s identity-verification workflow. What the ATS documentation does establish is that ATS and TVS are deeply integrated in CBP's biometric screening architecture, which Mobile Fortify appears to rely on. Cross-linking records helps, but it does not close the gap on its own.

Investigative reporting surfaced a February 2025 Privacy Threshold Analysis (PTA) that explains why no standalone Mobile Fortify SORN or PIA exists. CBP and ICE concluded that existing documentation provides sufficient coverage, despite acknowledging that Mobile Fortify is privacy-sensitive and requires both a PIA and a SORN \cite{Cox2025}. The absence of a Mobile Fortify–specific notice is a deliberate determination, not an administrative oversight. We have no comparable reporting regarding the seeming lack of an ICR filing.
The disclosure record is both more available and less complete than it first appears. The AI Use Case Inventory names the application and describes two deployed, high-impact use cases;  but its risk management fields are listed as in progress, and associated systems are identified only by opaque shorthand. Moreover, no information was publicly accessible during the months Mobile Fortify was operational before the January 2026 publishing of the Inventory. In addition to this time lag, we also identify that this deployed, high-impact system is operating while likely remaining non-compliant with its oversight requirements and vague in its details of integration with other government AI infrastructure.

\section{Structural Limits of Cross-Regulatory Comparison}

\subsection{Limited direct visibility of AI systems}
Searching the corpus of 3,735 SORNs for terms drawn from our AI search configuration,  including “artificial intelligence," “machine learning," “algorithm," and “automated decision" yielded few results. The term "artificial intelligence" appeared in only 6 SORNs; ``machine learning" in 4. Broader proxy terms fared better: ``algorithm" or "algorithmic" matched 34 SORNs, ``risk assessment" matched 90, and ``automated targeting" matched 30, for a total of 227 unique SORNs (6.1\% of the corpus) containing at least one term from the search configuration. A zero-shot classification pass using a LLM identified 554 SORNs (14.8\%) as AI-related when given access to full document text;  a substantially higher figure., 
AI-related systems are documented in SORNs but without the vocabulary of AI disclosure. 

Cross-referencing the 2023-2025 AI Use Case Inventories (710, 2,133 and 3,542 records, respectively) with the SORN corpus makes the divergence concrete: some AI inventory use cases describe systems that collect or process personal information and therefore should, under the Privacy Act, be covered by a SORN  yet those SORNs describe the underlying record system without identifying it as AI-driven. 20 AI inventory records could be matched to a corresponding SORN; of those, none contain AI-related terminology.

The ICR corpus tells a similar same story. Our extraction pipeline covers 37,871 ICRs filed with OIRA from January 2018 through March 2026, with 171,560 supporting documents converted to searchable text. Even with this full-text infrastructure in place, a two-stage keyword filter;  first identifying documents structured as OMB supporting statements, then filtering for any term from the AI search configuration, yielded only 1,160 candidate documents. LLM classification of the candidate set identified 1,144 ICRs as AI-related at high or medium confidence, representing a small percentage of the full corpus. As with SORNs, the dominant matched terms were functional descriptors: ``risk assessment," ``automated targeting," ``algorithm", rather than the language of AI systems.
The terms ``artificial intelligence" and ``machine learning" appeared rarely in supporting statements even for systems the AI Use Case Inventory explicitly labels as AI-driven, indicating that agencies filing ICR paperwork are not yet using the vocabulary of AI transparency, and that ICR disclosures will systematically undercount AI-adjacent systems under keyword-based searches. 

Taken together, the low hit rates across all three corpora using explicit AI terminology reflect a structural feature of federal disclosure practice rather than an absence of AI systems: each regime has its own vocabulary, shaped by the legal mandate it serves. SORNs describe record systems under Privacy Act authority; ICR supporting statements justify information collection burden under the Paperwork Reduction Act; 
only the AI Use Case Inventory is organized around AI.
The result is that even a researcher with full-text access to all three corpora, searching for AI by name, would substantially undercount the AI systems visible in the inventory.

\subsection{Limited overlap across transparency regimes}

Cross-regime analysis of explicit linkages reveals that government AI disclosures remain largely siloed, with minimal cross-references across the SORN, ICR, and AI Use Case Inventory frameworks. Of 554 AI-related SORNs, only 14 (2.9\%) appear anywhere in the ICR corpus. Restricting the comparison to ICR filings that are themselves AI-related reduces that figure further, to 8 SORNs (1.6\%). The remaining 97\% of AI-related SORNs exist in isolation from any ICR filing. The AI Use Case Inventory fares no better: of all entries in the inventory, only two reference a SORN; both belonging to DHS's Automated Targeting System. After cross-document resolution, we find that between 80-97\% of the documents (80\% for SORNs, 97\% for AI Use Case Inventory, and 82\% for ICR records) contain no reference to another document (see Appendix Figure~\ref{fig:link_distribution} for more details).

The near-absence of cross-references does not, on its own, resolve whether the same systems appear across regimes. 
To recover implicit connections, we applied a multi-stage linking pipeline to the three AI-related corpora (554 SORNs, 1,146 ICR filings, and 5,268 AI inventory entries) 
to classify candidate pairs as the same underlying system, connected by a data flow or shared component, or unlinked. Even under this permissive standard, which does not require that agencies have acknowledged any connection, the rate of confirmed cross-regime matches is low: 446 pairs were classified as the same system across at least two regimes, representing 20\% of AI-related records in the smaller of each paired corpus. A further 15,000 pairs were classified as sharing a data flow or common component---systems disclosed separately whose filings nonetheless describe a traceable dependency---a category that itself illustrates the fragmentation: a single capability may generate distinct records in each regime with no document pointing to the others.

\subsection{The AI Use Case Inventory as an instructive tool}

The federal AI Use Case Inventory is now in its third cycle. The 2023 inventory comprised 710 entries; by 2025 the program had grown to thousands of disclosures. Across the 2024 and 2025 inventories, 1,227 use cases appeared in 2024, 2,247 appeared for the first time in 2025, and 823 present in 2024 were absent from 2025 (Appendix Figure~\ref{fig:longitudinal_summary}). Because neither inventory discloses persistent identifiers, year-over-year continuity was established via exact and fuzzy matching (see Appendix~\ref{app:inventory_processing}). The volume of  new entries reflects continued expansion of federal AI activity. The 823 disappearances are harder to interpret; they may indicate the conclusion of operation of certain AI systems, reclassification, or something else entirely.
Science and Administrative Functions account for the largest share of use cases, but high-impact designations are concentrated in Law Enforcement (which includes immigration enforcement) and Health \& Medical 
(Appendix Figure~\ref{fig:topic_areas}). Of course, volume is an inadequate proxy for risk.

\begin{figure}
    \centering
    \includegraphics[width=\linewidth]{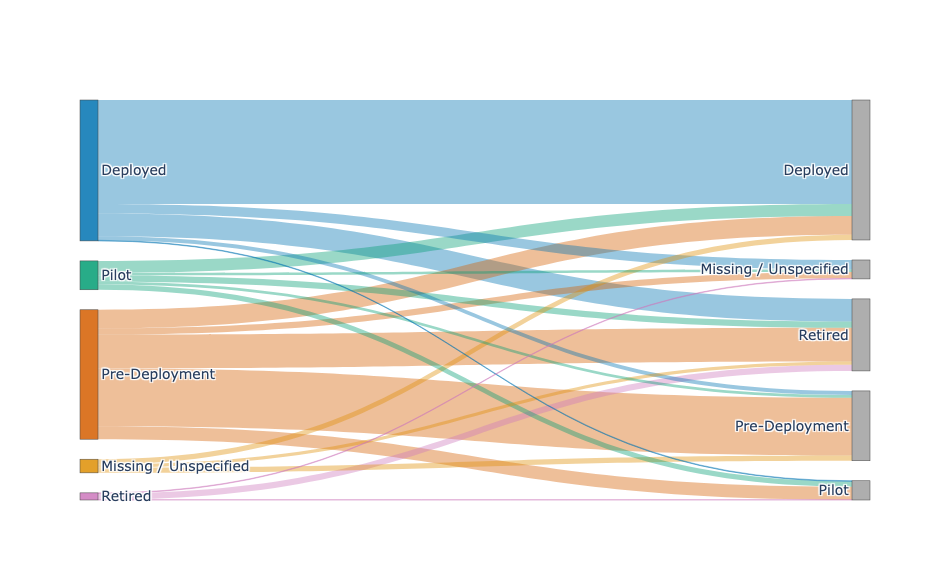}
    \caption{Sankey diagram demonstrating how the stages of development of the 1,227 AI use cases that were referenced in both the 2024 and 2025 Inventory changed. Most deployed AI technology continues to be in use in 2025, but heterogeneous outcomes occur.}
    \label{fig:sankey_all}
\end{figure}

Among the 1,227 use cases traceable across both inventories, most deployed technology from 2024 remains active in 2025, but outcomes are not uniform (Figure~\ref{fig:sankey_all}). Pre-deployment systems most commonly remain in pre-deployment after a year. Among systems that did change stage, the most frequent transition is retirement. The third most common outcome is advancement: pre-deployment systems that are now deployed.

Considering the high-impact use cases and their required risk management disclosures, we find that a number of agencies and use cases fail to fill out the required fields. For example, there are 14 use cases that don't have an appeals process in place, and at least two use cases that have failed to respond to whether they have an AI impact assessment completed. In addition, qualitative review of the data shows that others have labeled required fields for high-impact deployed use cases with ``not applicable'' without any additional justification and many still report that prerequisite impact assessment and risk management requirements are ``in progress'' rather than completed for already deployed systems.

\section{Policy Recommendations}

In considering how to make government transparency disclosures related to AI more effective and useful, we must also consider the audience and associated goals of such disclosures. Here, we are interested in disclosures of AI systems for a public audience with the goal of transparency into high-impact government operations. We thus focus our recommendations on improvements to the AI Use Case Inventory, as the existing mechanism that most closely matches this audience and goal. We note that, unfortunately, the AI use case inventory is currently set to no longer be required by statute as of 2027 \cite{advancingAIact2022}, although the Executive Order on Trustworthy AI \cite{eo13960} may still be in force beyond that date. Thus, all these recommendations stem from our first key recommendation: \textbf{continue conducting the AI use case inventory in perpetuity.} While these recommendations are described in the context of this U.S. based transparency mechanism, they have broader implications for AI transparency. 

\paragraph{Ensure clarity and broad applicability of the covered AI definition} In the Inventories from 2023, 2024, and 2025 we see the results of three different variants of AI definitions and associated guidance. While all three years used the same core AI system definition, the 2024 inventory added clarifications to the definition (e.g., clarifying that it included small systems and basic machine learning tools like logistic regression) \cite{omb_m2410}. These clarifications were effective and drastically increased the number of systems reported (from 710 disclosed in 2023 to 2,133 in 2024). The 2025 inventory however appears to have included agency interpretations of the high-impact AI system definition that relied heavily on the part of the definition that limits such systems to those ``whose output serves as a principal basis for a decision or action.'' The ``principal basis'' language served to remove items from the inventory that under the 2024 interpretation were presumed included and appeared in the inventory. An AI use case can often serve as \emph{a} basis for a decision or action, while questionably meeting the threshold of being a ``principal basis.'' Broad applicability of such definitions is important for public insight into these tools, and for ensuring high-impact systems are covered by the required risk management steps; a simple fix in this case would remove the principal basis clause.

\paragraph{Create consistent AI system identifiers and cross-reference with other disclosure regimes} The AI use case inventory is an annual disclosure regime and agencies are expected to report out on use cases each year, including on use cases that they've disclosed in a previous annual report. In such cases, it would be helpful to the public's ability to track changes in use cases over time to have a common identifier per AI system that is included in each year's inventory. An identifier could also include the possibility of versioning via a main key for the AI system based on the use case and an associated version number, with directions given to agencies by OMB as to what constitutes a new version of the AI system. Such identifiers should also be referenced, as appropriate, in other disclosures, for example future SORNs could be required to list the AI use case identifiers for any AI systems expected to make use of the data. The creation of such identifiers is standard practice for other similar disclosure regimes, including SORNs and ICRs. 
Future AI inventories should also directly link to associated SORNs and ICRs. 
Creating links between these regimes would usefully allow the public a view into the full system that includes the AI system description, the data it relies on, and any previous privacy-related disclosures. 

\begin{figure}
     \centering
     \includegraphics[width=\linewidth]{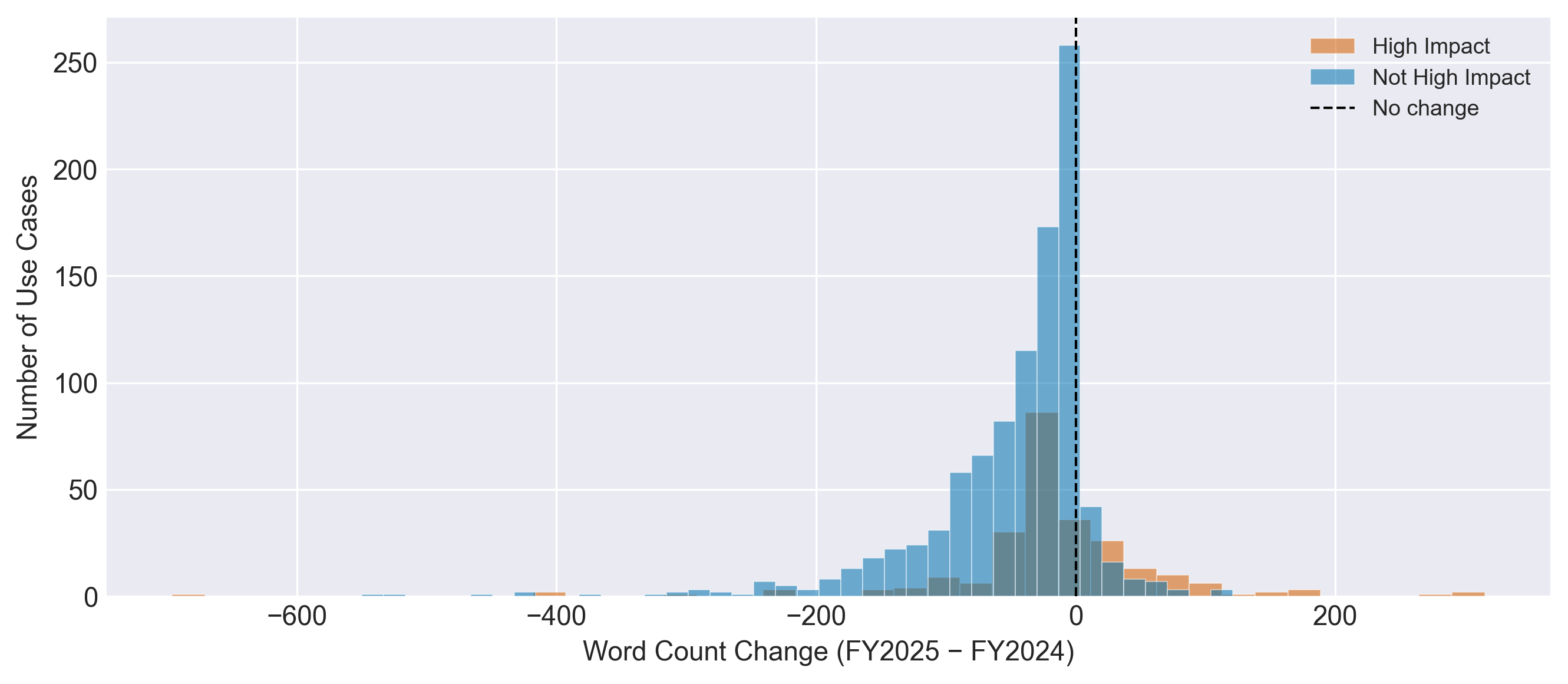}
     \caption{Overall we note a trend towards shorter responses in the AI use case entries that existed in both 2024 and 2025.}
     \label{fig:word_count}
 \end{figure}

\paragraph{Provide public transparency into risk management processes} In response to questions such as, ``What are the key risks from using the AI for this particular use case and how were they identified?" the 2024 Inventory included some thoughtful descriptive examinations of potential risks and benefits of AI uses cases, although not all agencies released thorough assessments. Some by the Department of Veterans Affairs (VA) were notable for their nuanced and careful examination of risks and prevention of potential harms. For example, the VA created a risk assessment algorithm for opioid patients (Stratification Tool for Opioid Risk Mitigation, or STORM). In the 2024 inventory, key risks are detailed including an over-reliance by providers on the tool, and mitigations such as provider training are described; the full response is 367 words and identifies and responds to seven risks. Unfortunately, while the 2025 Inventory includes a question to agencies asking for potential impacts, no responses to that question were released publicly. This is a trend towards decreased length in responses across agencies visible in the number of words in the Inventory (see Figure~\ref{fig:word_count}).
It's possible that agencies did still submit these more nuanced risk descriptions to OMB, but that OMB chose not to release those to the public this year. The public should have insight into these key risks and management processes.

\paragraph{Tailor inventory questions to the goal of transparency into high-impact use cases} While transparency into risk management assessments and practices should be reinstated in the Inventory, these were only ever required for high-impact use cases. This appropriately puts agency reporting effort on cases already identified as high-impact. Other responses may be unnecessary for use cases that are not high-impact, or there may be ways to streamline agency responses so that they are both more useful and easier for the agency. For example, linking to SORNs and ICRs could help agencies fully describe underlying data used by a system without further paperwork. Other changes could be made via user testing the Inventory's questions with agencies themselves.

\paragraph{Make the inventory more publicly usable and archived}
In our view, a key goal of the AI use case inventory and the associated improvements we suggest here is public transparency. While the SORNs, ICRs, and AI use case inventory are technically public disclosures, they are not made available in a way that is usability tested and easy for the public to browse and interact with. SORNs and ICRs are released via the Federal Register, which is a fully text-based format and has the advantage of existing archival processes. The AI use case inventory is released on government websites, some of which have already disappeared in the changeover between the Biden and Trump Administrations, and on GitHub in CSV and Excel formats. Releasing the inventory with clearly comparable fields has allowed journalists following government technology changes to analyze the inventory and successfully provided some public transparency through their reporting (see, e.g., \cite{fedscoop2025inventory}). However, the inventory should be maintained in an archival form and released via a website of the type we create (see Section \ref{sec:website}) or other easily accessible format that allows searching by field or keyword, and which is user-tested with the general public.

\section{Conclusion}
Given that the AI Use Case Inventory is still relatively new, we should take advantage of the associated interpretive flexibility and continue to make improvements to make the disclosures more meaningful to the public. We describe how triangulating AI disclosures across different transparency regimes reveals more about AI systems despite structural limitations, and make policy recommendations to improve existing transparency regimes.

\section*{Acknowledgements}This work was supported in part by a grant from the Ford Foundation.

\bibliography{govaifiles}
\appendix
\clearpage
\section{Appendix}
\subsection{Document Processing Details}

\subsubsection{ICR Data Processing}
\label{app:icr_processing}
We collected 37,871 ICR records submitted to OIRA between January 2018 and March 2026. Of these, 31,670 (83.6\%) had supporting documents successfully downloaded to per-ICR folders on disk. The remaining records either failed during extraction.  Across the 31,670 records with downloaded documents, we obtained 253,304 files in total, with a median of 4 files per ICR and a mean of 8 files. We then converted all PDFs and DOCX files to plain text yielding 171,560 markdown files that form the corpus for keyword filtering and LLM classification.

Because of the scale of the documents, we applied a two-stage keyword filter to identify PRA documents likely to describe AI-related information collections before submitting them to the LLM classifier. This follows the approach outlined in similar cross-document co-reference to narrow the candidate space before doing classification tasks.

We wanted to surface documents including supporting statements, Privacy Impact Assessments (PIAs), Privacy Threshold Analyses (PTAs), Systems of Records Notices (SORNs), Algorithmic Impact Assessments (AIAs), and civil liberties assessments. A document was included for LLM classification if either the filename matched regular expression patterns for these document types or the content included document type-specific structural headers (with a match being considered $\geq$ 2 text headers matching known sections of those documents).

Both the filename patterns and the structural header sets were developed by inspecting documents from across the corpus; they reflect the naming conventions and section structures that agencies use in practice rather than a formally specified taxonomy, and are not guaranteed to be exhaustive.

In the second stage, surviving documents were screened for a vocabulary of 22 AI-relevant terms drawn from a shared search configuration used across all three disclosure regimes. The vocabulary was developed iteratively through manual review of documents: terms were added when recurring language in agency filings was not yet captured, and the list was refined until successive reviews produced diminishing new high-quality matches. The resulting vocabulary is heuristic rather than exhaustive. It reflects the terminology that federal agencies use in practice when describing these systems, which does not always align with academic or technical conventions. Terms were grouped into five conceptual categories:

\begin{itemize}
  \item \textit{Core AI and ML terminology}: artificial intelligence, machine learning, deep learning, neural network, algorithmic, algorithm, predictive
  \item \textit{Automated processing and risk scoring}: automated targeting, automated decision, risk assessment, risk score, risk scoring
  \item \textit{Rule-based systems}: rule-based, rule-managed
  \item \textit{Computer vision and biometrics}: video analytics, image analysis, object detection, license plate
  \item \textit{Computer matching}: computer match, computer matching,
    automated matching, data matching
\end{itemize}

\noindent All terms were matched case-insensitively against full document text. The computer matching category warrants particular note: Privacy Act computer matching programs describe automated cross-dataset comparisons that may not employ machine learning but nonetheless constitute algorithmic decision-making subject to federal oversight requirements. Including this category ensures coverage of a disclosure regime---the Computer Matching and Privacy Protection
Act---that intersects with but predates contemporary AI policy. A document was retained as a candidate if any term matched. This two-stage filter produced 5,670 candidate documents across 3,378 ICRs.

\subsubsection{SORN Data Processing}
SORNs were standard to process, they had consistent field names and were accessible via the Federal Register API. 

We identified AI-relevant Systems of Records Notices (SORNs) using a zero-shot large language model classification workflow applied to a corpus of federal SORN JSON files.The underlying corpus consisted of 3,735 SORNs spanning 2010-2025 pulled from the Federal Register API. %
We then classified these SORNs as AI-related using zero-shot classification on GPT-5.4 mini. We randomly sample 122 SORNs predicted to be AI-related and, and find that the classification had 69.7\% accuracy.

AI relevance is not reported directly in any standard SORN field; instead, it had to be inferred from the notice’s description of the system’s functions, which we do through LLM prompts. The authors engaged in a preliminary review of results, and then made tweaks to the prompt to better account for the standardized structure of the SORN. The final prompt, available in Appendix~\ref{app:sorn-prompt}, distinguished systems that perform matching, scoring, recognition, prediction, ranking, approval or denial support, risk assessment, or other automated analytical judgments from systems that primarily store, retrieve, exchange, or display records. It also returned a binary \texttt{is-ai-related} indicator, a flag for whether machine learning or AI was explicitly stated in the notice, short quoted evidence phrases copied from the SORN text, and brief reasoning. Model outputs were then manually reviewed as part of the evaluation process.

\subsubsection{AI Use Case Inventory Processing}
\label{app:inventory_processing}
Federal agencies submitted their AI use case inventories using the OMB-provided template, but column names in submitted files vary substantially across agencies and across the 2023, 2024, and 2025 Inventory years. We standardized all three years into a 45-field canonical schema using a pipeline built around per-agency
JSON mapping specifications. The schema defines a shared set of standard column mappings from OMB template question strings to short canonical field names, which each per-agency spec can override to handle agency-specific column name variants. Of the 26 per-agency 2025 specs, 20 agencies used the standard OMB template without modification; the remaining 6 required full custom column mappings. Unmapped source columns are serialized to a JSON \texttt{agency\_extensions}
field on each record rather than discarded.

Controlled-vocabulary fields required additional normalization across years, as the OMB template encodes answer choices as verbose prefix-lettered strings that changed wording between reporting cycles. The pipeline resolves these to consistent canonical values for \texttt{stage\_of\_development}, \texttt{high\_impact\_status}, \texttt{development\_source\_type}, and the nine high-impact risk management practice fields, using per-field lookup tables with
prefix pattern matching as a fallback. Where the 2024 source lacked a
\texttt{high\_impact\_status} column, that field is derived from the
\texttt{rights\_safety\_impact\_type} field per OMB M-24-10 definitions.
Unparseable values are preserved in \texttt{agency\_extensions} and flagged for manual review.

\subsubsection{Longitudinal AI Use Case Inventory Tracking}
\label{app:inventory_longitudinal}
Given that there are no persistent identifiers across years for the AI Use Case Inventory entries, we adopt the following matching procedure. We linked records pairwise (2023--2024 and 2024--2025) using a two-pass procedure. In the first pass, a record in year $t$ is matched to a record in year $t+1$ if the composite key (\texttt{use\_case\_name}, \texttt{canonical\_abbreviation}) is identical. Unmatched records then proceed to a fuzzy pass, restricted to pairs sharing the same \texttt{canonical\_abbreviation} to reduce false positives across agencies. For each within-agency candidate pair, we compute a composite similarity score as $0.65 \times s_{\text{name}} + 0.35 \times s_{\text{purpose}}$, where both components use \texttt{rapidfuzz} token-sort ratio; when either record is missing a \texttt{purpose\_and\_benefits} value, the score falls back to name similarity alone. The match threshold is 65 for the composite score and 70 for the name-only fallback (based on manual review of matches on threshold). Each year-$t+1$ record is assigned at most one match, resolved greedily by descending score. For 2023--2024, this yielded 183 exact and 36 fuzzy matches, with 415 records classified as dropped and 1,859 as new. For 2024--2025, 1,044 exact and 179 fuzzy matches were found, with 833 records dropped and 2,251 new. Of the 710 use cases reported in 2023, 104 were matched continuously through all three years.

\subsection{Classification Prompts}

\subsubsection{SORN Classification Prompt}
\label{app:sorn-prompt}

\noindent System of Records Notices (SORNs) were classified using \texttt{gpt-5.4-mini}
via the OpenAI Batch API. The system prompt is reproduced in full below.

\begin{quote}\small\noindent
\textit{System role:}

You are a U.S.\ federal government records expert analyzing System of Records Notices
(SORNs) to determine whether the described system is AI-related or uses algorithmic
decision support.

Return exactly one JSON object with these fields:
\begin{itemize}[noitemsep,topsep=2pt]
  \item \texttt{classification}: one of \texttt{automated\_decision\_support},
        \texttt{biometric\_matching}, \texttt{automated\_recognition},
        \texttt{recordkeeping\_only}, \texttt{unclear}
  \item \texttt{is\_ai\_related}: boolean
  \item \texttt{ml\_explicitly\_stated}: boolean
  \item \texttt{confidence}: \texttt{high}, \texttt{medium}, or \texttt{low}
  \item \texttt{evidence}: array of 1--3 phrases copied verbatim from the SORN
  \item \texttt{reasoning}: string (1--2 sentences)
\end{itemize}

\noindent\textit{Decision process} (applied in order):
\begin{enumerate}[noitemsep,topsep=2pt]
  \item Determine whether the SORN says the system itself performs matching, scoring,
        recognition, prediction, ranking, approval/denial support, risk assessment, or
        other automated analytical judgments about people, vehicles, or goods.
  \item Classify the system based on its own function, not on connected systems that
        only share data with it.
  \item If the system only stores, retrieves, replicates, exchanges, or displays
        records, classify it as \texttt{recordkeeping\_only}.
  \item If the text is genuinely ambiguous, classify it as \texttt{unclear} instead
        of guessing.
\end{enumerate}

\noindent\textit{Classification rules:}
\begin{itemize}[noitemsep,topsep=2pt]
  \item \textbf{automated\_decision\_support}: The system performs rules-based
        targeting, screening, risk scoring, anomaly detection, watchlist matching, or
        other automated decision support, including cases where ML may or may not be
        explicitly stated.
  \item \textbf{biometric\_matching}: The system performs face, fingerprint, iris,
        voice, or similar matching, identity verification, liveness detection, or
        one-to-one/one-to-many comparison.
  \item \textbf{automated\_recognition}: The system performs automated license plate
        recognition or comparable automated recognition of observed entities.
  \item \textbf{recordkeeping\_only}: The system is primarily a database, records
        repository, case management system, ingestion layer, or evidence store without
        a described matching, scoring, recognition, or predictive function.
  \item \textbf{unclear}: The SORN suggests possible algorithmic use, but the
        system's own function is not described clearly enough to classify confidently.
\end{itemize}

\noindent\textit{Constraints:} A system is not marked AI-related solely because it
mentions databases, storage, cloud infrastructure, surveillance cameras, social media,
geolocation, or ``new technology.'' Facial matching, liveness detection, fingerprint
and voice recognition, and automated license plate recognition are always treated as
AI-related. Rules-based targeting and automated risk scoring count as AI-related even
without explicit ML. The primary classification signals are the \texttt{Purpose} and
\texttt{categories\_of\_records} fields; among secondary signals, \texttt{routine\_uses}
is strongest, as it often contains explicit language about automated referral, scoring,
profiling, or algorithmic targeting.
\end{quote}

\subsubsection{ICR Prompt:}
\label{app:pra-prompt}

\noindent PRA supporting statements were classified using \texttt{gpt-5.4-mini}
via the OpenAI Batch API. The system prompt is reproduced in full below.

\begin{quote}\small\noindent
\textit{System role:}

You are analyzing federal PRA (Paperwork Reduction Act) supporting statements to
determine whether the described information collection involves AI, machine learning,
automated decision-making, computer matching, or algorithmic processing.

Return exactly one JSON object with these fields:
\begin{itemize}[noitemsep,topsep=2pt]
  \item \texttt{is\_ai\_related}: boolean
  \item \texttt{confidence}: \texttt{high}, \texttt{medium}, or \texttt{low}
  \item \texttt{evidence}: array of 1--3 short phrases copied verbatim from the document
  \item \texttt{reasoning}: string (1--2 sentences)
  \item \texttt{system\_name}: full name of the federal AI system or program described,
        exactly as written in the document (\texttt{null} if not explicitly named)
\end{itemize}

\noindent Mark \texttt{is\_ai\_related} true if the supporting statement describes
or mentions:
\begin{itemize}[noitemsep,topsep=2pt]
  \item AI, machine learning, neural networks, computer vision, or predictive models
  \item Automated targeting, risk scoring, or anomaly detection
  \item Computer matching or automated data matching across datasets
  \item Algorithmic processing of applications, claims, or records
  \item Biometric processing (facial recognition, fingerprint matching, etc.)
  \item Automated decision support or rules-based screening
\end{itemize}

\noindent Mark \texttt{is\_ai\_related} false if the collection is solely about
gathering information from the public, recordkeeping, or standard survey/form
administration with no described AI or automated analytical processing.

Evidence must be short phrases copied verbatim from the document, not paraphrases.
If no clear evidence exists, set \texttt{is\_ai\_related} to false.
\end{quote}

\subsection{LLM Reranker Prompt}
\label{app:reranker_prompt}
\textit{System role:} You are an expert in U.S. federal government AI governance and regulatory filings. Your task is to classify the relationship between two regulatory documents --- whether they refer to the same underlying AI system, to distinct systems connected by a data flow or shared component, or to unrelated systems. Records may use different terminology, be filed by different offices, or describe the same system at different levels of detail. Distinct systems that share a data flow, a common component, or infrastructure should be marked \texttt{related}, not \texttt{link}.

\textit{User role:} Below are two regulatory records from different U.S. federal transparency regimes. Classify the relationship between them by working through the questions below before giving your final answer.

\medskip
\noindent RECORD A\\
\textit{[Regime, Record ID, Agency, Bureau/component, System/title, Date, Description]}

\medskip
\noindent RECORD B\\
\textit{[Regime, Record ID, Agency, Bureau/component, System/title, Date, Description]}

\medskip
\noindent Work through each question in order:
\begin{enumerate}[noitemsep,topsep=2pt]
\item \textbf{Name and identifier match:} Do the system names, official identifiers (e.g., SORN number, system ID such as DHS/CBP-022), or acronyms identify the same system --- accounting for common aliases and abbreviations?
\item \textbf{Subject and function:} Is each record's central system the same entity performing the same function? Or are they distinct systems connected by a data flow or shared component (e.g., System A ingests records produced by System B, or both rely on the same underlying data pipeline)?
\item \textbf{Explainable discrepancies:} If names, agencies, or purposes differ, is there a plausible explanation --- cross-component use, a bundled filing covering multiple related forms, or different regulatory angles on the same underlying system?
\end{enumerate}

\noindent Return exactly one JSON object with these fields:
\begin{itemize}[noitemsep,topsep=2pt]
\item \texttt{decision}: \texttt{link}, \texttt{related}, \texttt{uncertain}, or \texttt{no\_link}
\item \texttt{confidence}: \texttt{high}, \texttt{medium}, or \texttt{low}
\item \texttt{evidence}: array of short phrases copied verbatim from each record
\item \texttt{explanation}: string (one paragraph summarizing reasoning across all three questions)
\item \texttt{flags}: array of strings noting concerns (e.g., \texttt{same system name but different agencies}, \texttt{possible mission creep})
\end{itemize}

\noindent Decision definitions:
\begin{itemize}[noitemsep,topsep=2pt]
\item \texttt{link}: the two records refer to the same underlying AI system (different names, sub-agencies, or levels of detail are acceptable)
\item \texttt{related}: distinct systems connected by a data flow, shared component, or shared infrastructure
\item \texttt{uncertain}: cannot determine from these records alone
\item \texttt{no\_link}: different systems with no documented relationship
\end{itemize}

\noindent AI Use Case Inventory entries are typically brief summaries (100--200 words) and often omit system identifiers and data sources. Brevity in one record should not by itself push toward \texttt{uncertain} or \texttt{no\_link}. When evidence is mixed or one record is sparse, prefer \texttt{uncertain} over \texttt{no\_link} --- missing a true link is more costly than a false positive in this analysis.

\begin{figure*}
    \centering
    \includegraphics[width=\linewidth]{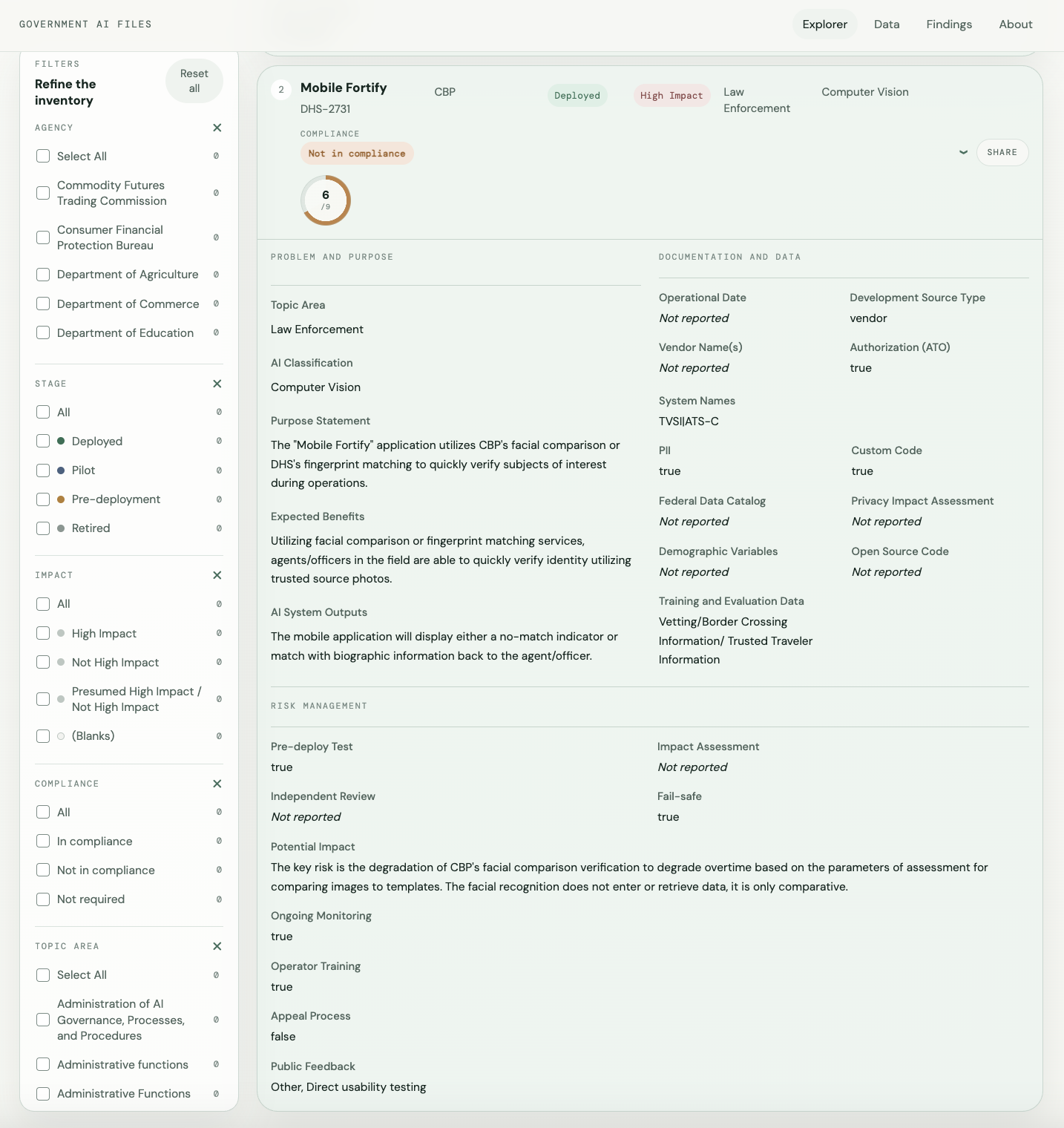}
    \caption{A screenshot of one of the two AI Use Case Inventory entries for Mobile Fortify on \website.}
    \label{fig:mobile_fortify}
\end{figure*}

\begin{figure}[htbp]
    \centering
    \includegraphics[width=\linewidth]{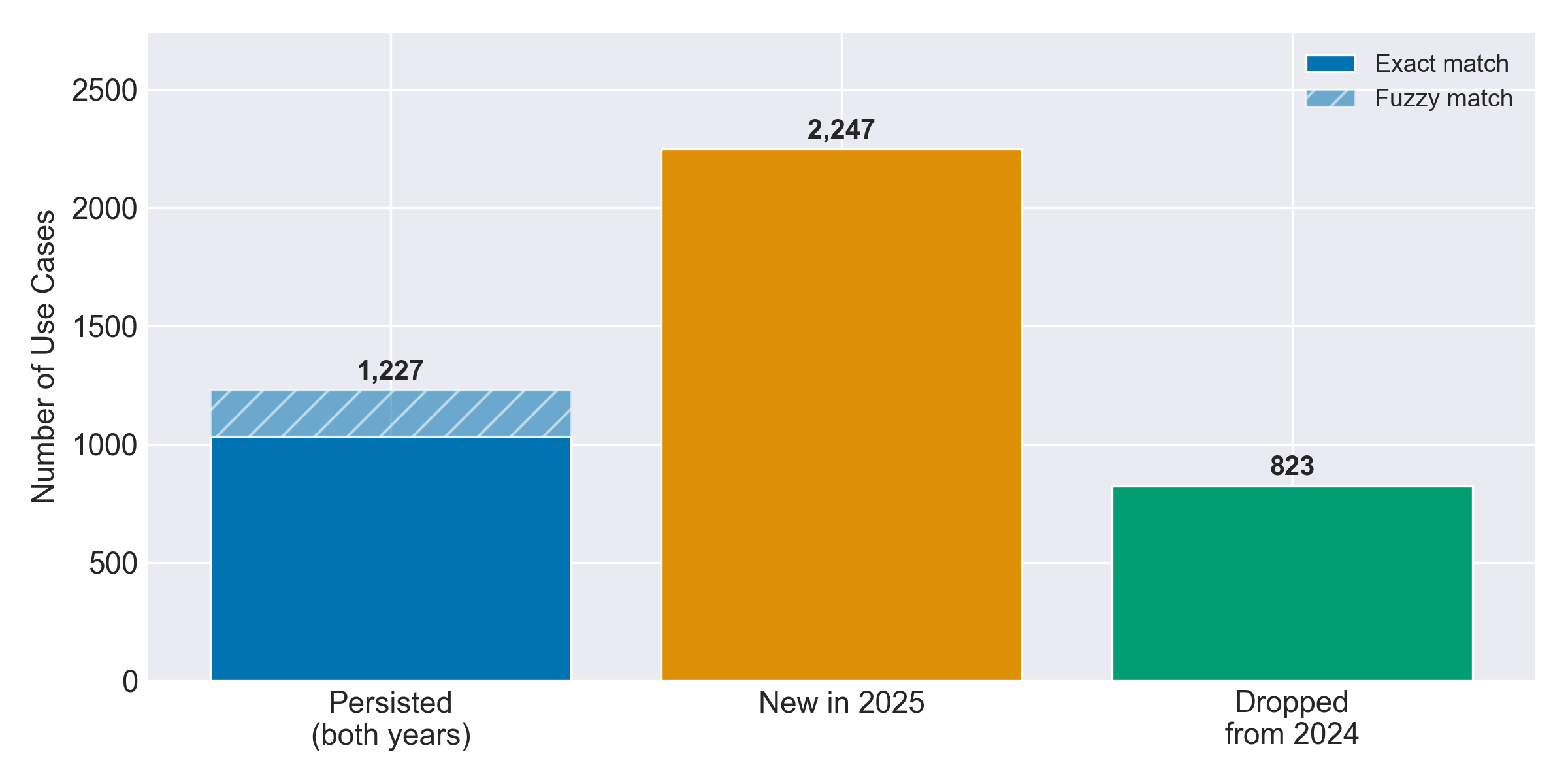}
    \caption{Longitudinal comparison of AI use cases between 2024 and 2025 inventory cycles. Bars show AI use cases persisting from 2024-2025 years (blue; N=1,227), first appearing in 2025 (orange; N=2,247), or absent from 2025 despite appearing in 2024 (green; N=823). Neither inventory discloses persistent identifiers; year-over-year continuity was established by name matching, with solid fill indicating exact matches and hatching indicating fuzzy matches.}
    \label{fig:longitudinal_summary}
\end{figure}

\begin{figure*}
    \centering
    \includegraphics[width=\linewidth]{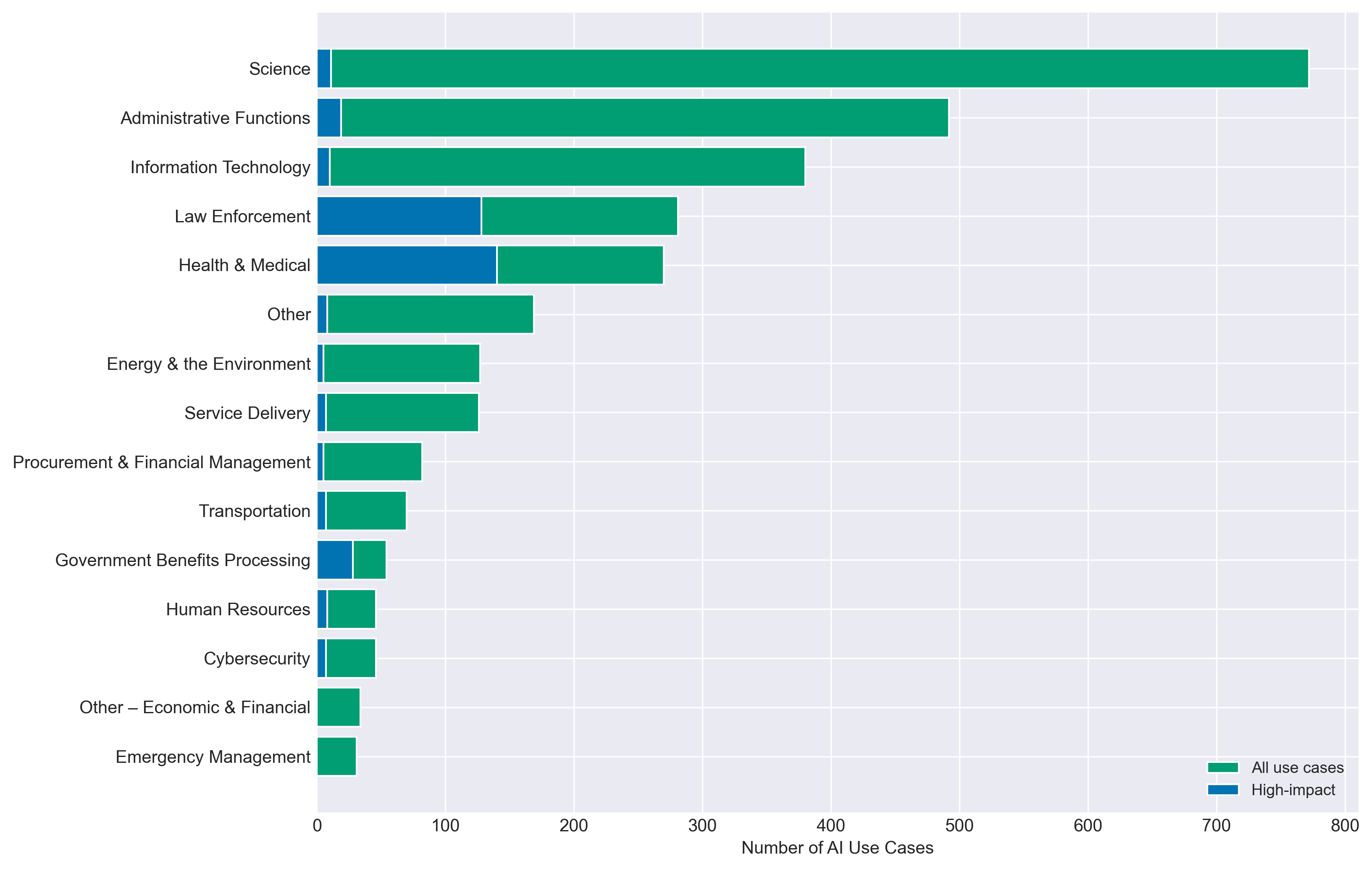}
    \caption{Bars show the number of use cases per topic area (green) alongside the subset designated high-impact (blue). Topic area classifications were revised between the 2024 and 2025 inventories, precluding direct year-over-year comparison by category. Science and Administrative Functions account for the largest share of use cases overall, but high-impact designations are concentrated in Law Enforcement and Health \& Medical.}
    \label{fig:topic_areas}
\end{figure*}

\begin{figure*}[htbp]
    \centering

    \begin{minipage}[b]{0.32\textwidth}
        \centering
        \includegraphics[width=\textwidth]{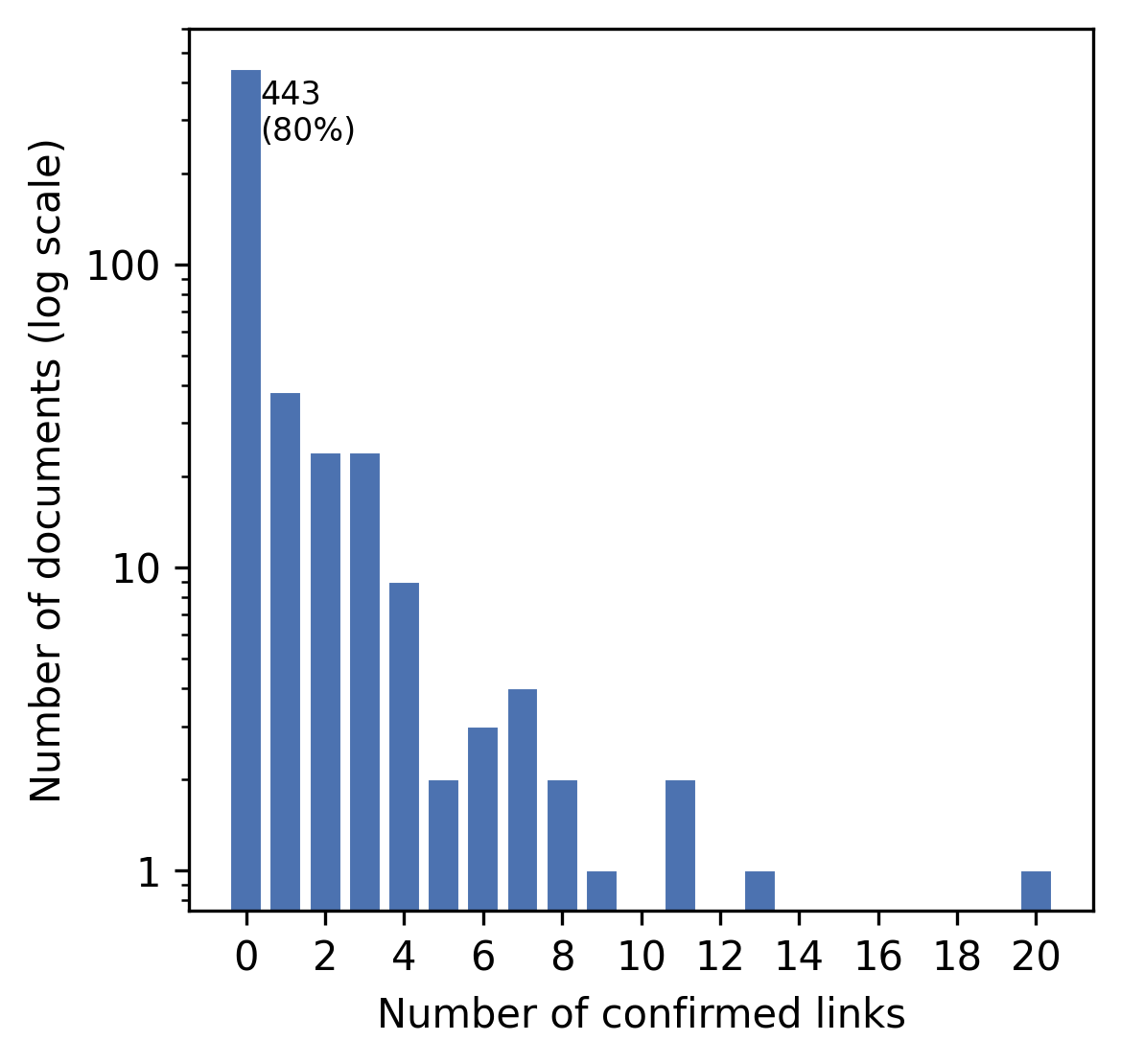}
        \small \textbf{(a)} AI-classified SORNs ($n=554$)
    \end{minipage}
    \hfill
    \begin{minipage}[b]{0.32\textwidth}
        \centering
        \includegraphics[width=\textwidth]{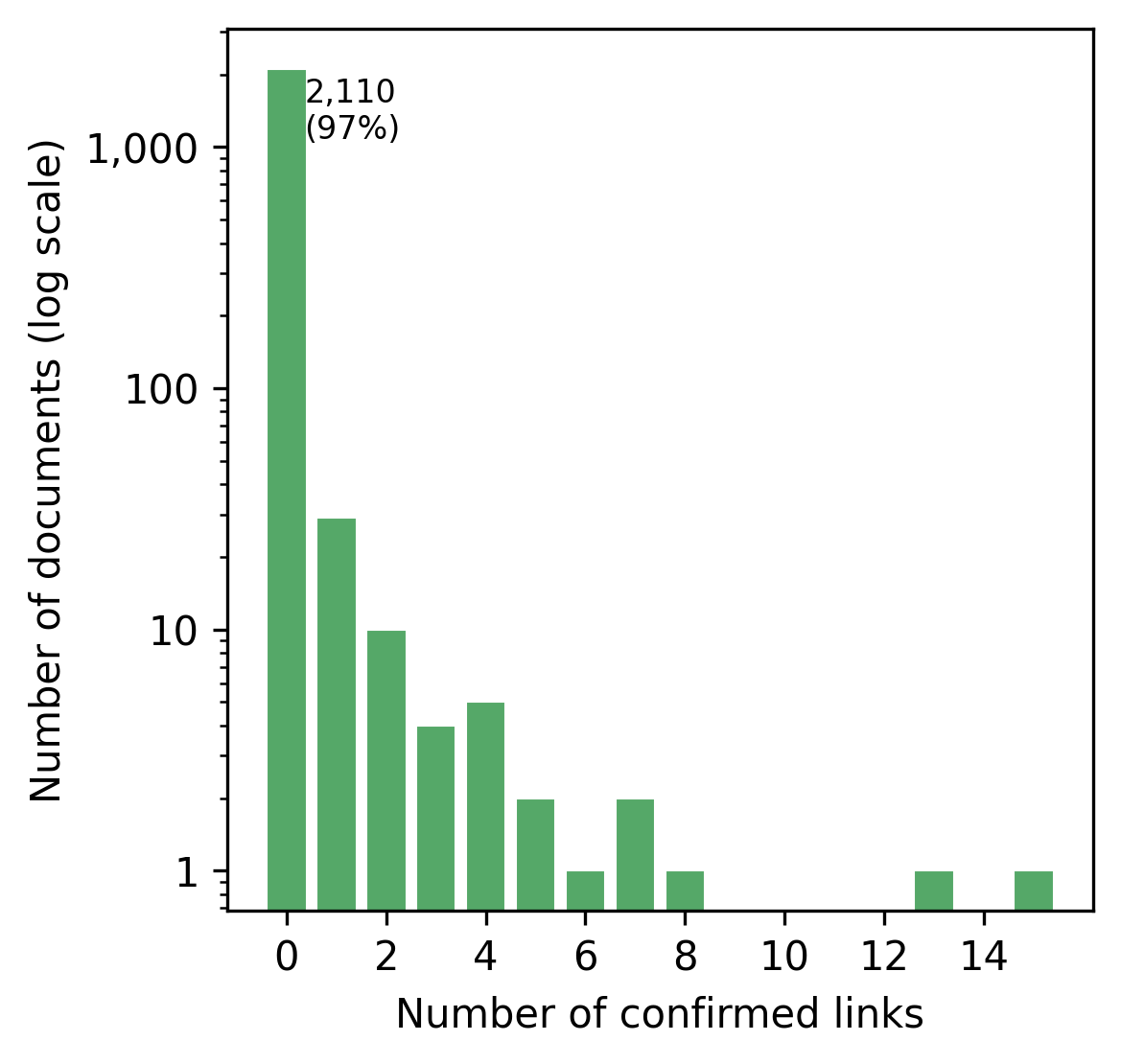}
        \small \textbf{(b)} AI Inventory records ($n=2166$)
    \end{minipage}
    \hfill
    \begin{minipage}[b]{0.32\textwidth}
        \centering
        \includegraphics[width=\textwidth]{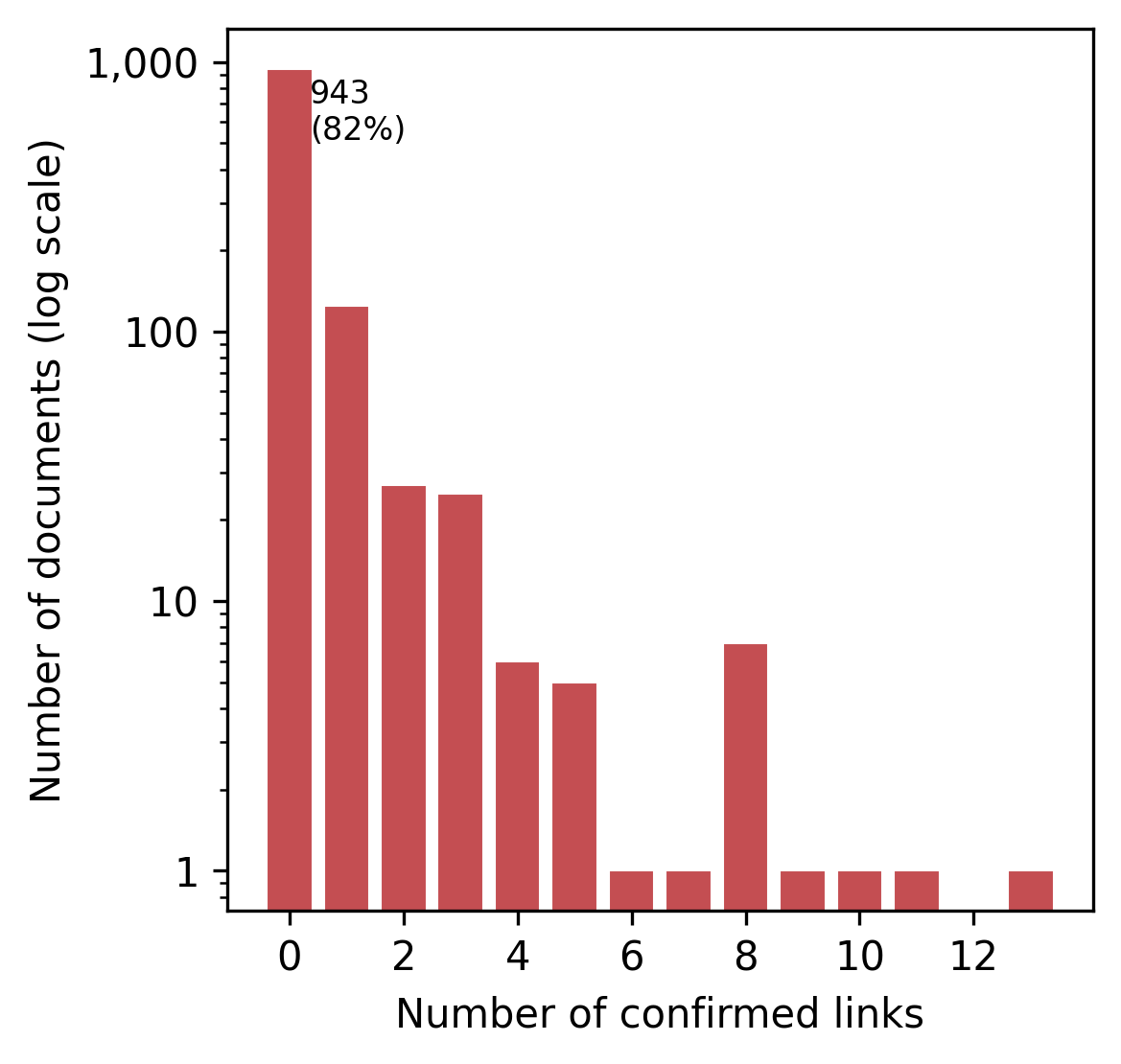}
        \small \textbf{(c)} AI-classified ICRs ($n=1144$)
    \end{minipage}

    \caption{Distribution of confirmed cross-regime links per document across three federal AI disclosure regimes. In panel (a), 443 SORN records (80\%) have no confirmed link; in panel (b), 2,110 AI Use Case Inventory records (97\%) have no confirmed link; and in panel (c), 943 ICR records (82\%) have no confirmed link    to a record in either of the other two regimes. $Y$-axes use log scale; only records appearing in the reranker candidate pool are included.}
    \label{fig:link_distribution}
\end{figure*}

\end{document}